\documentclass[twocolumn, superscriptaddress, aps, pra, reprint]{revtex4-2}
\usepackage[english]{babel}
\usepackage[utf8]{inputenc}
\usepackage{multirow}

\usepackage[normalem]{ulem}
\usepackage{amsmath}
\usepackage{url}

\usepackage{siunitx}
\usepackage[pdftex, pdftitle={Article}, pdfauthor={Author}]{hyperref}

\usepackage{graphicx}
\usepackage{dcolumn}
\usepackage{bm}

\begin{document}

\title{Interferometry with few photons} 

\author{Q. Pears Stefano}
\email{quimeymartin.pearsstefano@ehu.eus}
\altaffiliation{Present address: \emph{Centro de Física de Materiales, Paseo Manuel de Lardizabal 5, 20018 Donostia-San Sebastián, Spain}}
\affiliation{Universidad de Buenos Aires, Facultad de Ciencias Exactas y Naturales, Departamento de Física, Pabellón I, Ciudad Universitaria (1428), Buenos Aires, Argentina \\}
\affiliation{CONICET - Universidad de Buenos Aires, Buenos Aires, Argentina.\\}

\author{A. G. Magnoni}%
\email{magnoni.agustina@gmail.com}
\affiliation{Laboratorio de Óptica Cuántica, DEILAP, UNIDEF (CITEDEF-CONICET), Buenos Aires, Argentina}%
\affiliation{Universidad de Buenos Aires, Facultad de Ciencias Exactas y Naturales, Departamento de Física, Pabellón I, Ciudad Universitaria (1428), Buenos Aires, Argentina \\}

\author{D. Rodrigues}
\affiliation{Universidad de Buenos Aires, Facultad de Ciencias Exactas y Naturales, Departamento de Física, Pabellón I, Ciudad Universitaria (1428), Buenos Aires, Argentina \\}
\affiliation{CONICET - Universidad de Buenos Aires, Instituto de Física de Buenos Aires (IFIBA), Buenos Aires, Argentina.\\}

\author{J. Tiffenberg}
\affiliation{Fermi National Accelerator Laboratory, Batavia IL, United States}%

\author{C. Iemmi}%
\affiliation{Universidad de Buenos Aires, Facultad de Ciencias Exactas y Naturales, Departamento de Física, Pabellón I, Ciudad Universitaria (1428), Buenos Aires, Argentina \\}
\affiliation{CONICET - Universidad de Buenos Aires, Buenos Aires, Argentina.\\}

\date{\today}

\begin{abstract}

Optical phase determination is an important and established tool in diverse fields such as 
astronomy, biology, or quantum optics. There is increasing interest in using a lower 
number of total photons. However, different noise sources, such as electronic readout noise 
in the detector, and shot noise, hamper the phase estimation in regimes of very low illumination.
Here we report a study on how the quality of phase determination is affected by these two sources 
of noise. To that end, we experimentally reconstruct different wavefronts by means of a point 
diffraction interferometer for different mean intensities of illumination, up to $15\ \mathrm{phot/px}$. 
Our interferometer features a Skipper-CCD sensor, which allows us to reduce the readout 
noise arbitrarily, thus enabling us to separate the effect of these two sources of noise. 
For two cases of interest: a spatial qudit encoding phase, consisting of d = 6 uniform phase
 regions, and a more general continuous phase, we see that reducing the readout noise leads 
 to a clear improvement in the quality of reconstruction. This can be explained by a simple
  noise model that allows us to predict the expected fidelity of reconstruction and shows 
  excellent agreement with the measurements.
\end{abstract}


\maketitle

\section{Introduction}\label{sec:intro}

Interferometric measurements of light involve different techniques that allow to
know phase distributions with high precision. Although interferometry was
initially related to metrology of optical systems, its applications spread in
fields as diverse as astronomy, biology, quantum optics and many more. Current
advances achieved in computers and optoelectronic devices, such as spatial light
modulators and detectors, make interferometry a widely used tool in a large
number of applications \cite{Malacara, Schnars1994, Kim2010, Kacperski2006,
Zhang2019, PearsStefano2017}.
 
Digital recording and numerical processing of an interferogram led to
a great improvement in measurement techniques. It wasn't until the mid-1990s
that light sensors began to provide images of a certain quality. But, once the
initial objective was achieved, which was to increase their spatial resolution,
the challenge of achieving images with low light levels and good signal-to-noise
ratio (SNR) arose.

Two sources of noise corrupt the process of capturing an image: those that come
from the source (shot noise) and those that originate from the sensor (readout
noise). Even though increasing the signal (either the intensity or the exposure
time) seems an obvious way to improve the SNR, there are experiences where this
is not possible. Some alternative proposals include the use of unconventional
light sources such as squeezed states or entangled photons
\cite{Giovannetti2004,Mitchell2004a}. Unfortunately, these devices are not easy
to implement in most optical set-ups. Measurements at very low light levels have
applications in diverse fields including quantum information processing or
biological studies, where the sample can be damaged by a high photon flux. Some
of these experiences can be performed by using, for example, a highly attenuated
laser beam. In that scenario, a doubt arises: how many photons are needed to
calculate a phase distribution? In order to answer that question, it would be
necessary to be able to separate the noise caused by the light source
statistics, from the readout noise. A direct way of doing this is by eliminating
the latter using Skipper-CCD sensor \cite{Tiffenberg2017} which, by contrast
with other technologies, holds this feature in an unprecedented wide dynamic
range, from zero to a thousand photons per pixel \cite{Rodrigues2021}. In this
work, we propose to measure phase distributions at very low light levels in two
different situations: in one of them we perform quantum state tomography (QST)
on spatial q-dits (D-level quantum systems). In the other, we evaluate a an
arbitrary phase distribution: in this case we evaluate a continuous quadratic
wavefront. The phase reconstruction process is analyzed for different numbers of
photons per pixel and readout noise on the Skipper-CCD.

The article is organized as follows: in section \ref{sec:framework} we review the interferometric
process used to obtain the phase distribution, while in section \ref{sec:experimental} we describe
the experimental setup. In section \ref{sec:results}, on the one hand, we numerically simulate the
 expected results for different light intensities, taking into account the light source statistics and the
readout noise. On the other hand, the data acquisition method is explained and
experimental results are analyzed. Finally, in section \ref{sec:conclusions} we give the
conclusions.

\section{Methodology Framework}\label{sec:framework}
Phase-shifting interferometry (PSI) has proven to be an accurate and precise
method to evaluate wavefront phase distributions. In this technique, controlled
phase shifts are introduced between the reference and the tested beam
\cite{Creath1988}. The number of interferograms recorded as the phase is shifted
varies depending on the algorithm employed to recover the phase distribution of
the wavefront.

As it was previously mentioned, to diminish the readout noise, we used a
Skipper-CCD. Although this device requires a recording time similar to that of
conventional CCDs, the total acquisition time for each interferogram, including
the noise reduction process
($60\ \mathrm{\mu sec}/\mathrm{pixel}/\mathrm{sample})$, makes necessary the use
of a very stable interferometer. To this end, we implemented an architecture based on a point
diffraction interferometer (PDI) that was introduced by Linnik
\cite{Linnik1933}. Its common-path configuration, where the reference beam is
generated from the same wavefront under characterization, results on an
interferometer extremely stable against vibrations and air turbulence. When
combined with PSI schemes, for example by using liquid crystal technology
\cite{Mercer1996, Iemmi2003, Ramirez2013} to control the phase steps, the
potential applications of this interferometer are enhanced.

The reconstruction of the full wavefront $U(x,y)$ is done by using a the PSI architecture 
similar to that described in reference \cite{Iemmi2003}. Briefly, a convergent optical
 processor allows to apply at the Fourier Transform plane of the input wavefront 
 $U(x,y)$, a point-like phase filter. This element generates a diffracted beam 
 that is used as the interferometric reference. The resulting interferograms of 
 applying $N=4$ controlled phase shifts, $\alpha_n = 2\pi n/N,$ with 
 $n=0,\cdots,3$, are registered with the Skipper CCD.

The resulting amplitude for each of the phase shifts $\alpha_n$ at the image
plane $\Pi_i$ is
\begin{align}\label{eq:en}
  E_n(x,y) = U(x,y) + |K| e^{i\mu} \left[ e^{i \alpha_n} - 1\right],
\end{align}
where $|K|$ and $\mu$ are, respectively, the amplitude and phase of the
reference beam. It is worth noting that the reference beam corresponds to a
  plane wave with the mean value of $U(x,y)$. To highlight that the final aim
of this method is to obtain the the phase of $U(x,y)$, it is
useful to rewrite $U(x,y) = u(x,y) \exp(i \phi(x,y))$, where $u(x,y)$ is the
absolute value of $U(x,y)$ and $\phi(x,y)$ represents the phase. With the aid of
the following combination of all the interferograms
\begin{align}\label{eq:C}
  C(x,y) = \sum_{n=0}^{N} \left|E_n(x,y)\right|^ 2\cos\left(\frac{2\pi
  n}{N}\right)\\\label{eq:S}
  S(x,y) =  \sum_{n=0}^{N} \left|E_n(x,y)\right|^ 2\sin\left(\frac{2\pi n}{N}\right),
\end{align} the unknown phase of the wavefront $\phi(x,y)$ can be reconstructed
as
\begin{align} \label{eq:phase} \phi(x,y) = \mathrm{arctan2}(S, C-C_0) -
  \mu,
\end{align}
where $\mathrm{arctan2}(x_1, x_0)$ is defined as the angle between the
2-dimensional vector $(x_0 , x_1)$ and the $x_0$ axis, and
$C_0 = -N\left|K\right|^2$. The value of $C_0$ can be readily obtained from the
value of $C(x,y)$ at the points in which the input wavefront $U(x,y)$ is zero.
Finally, the real amplitude of $u(x,y)$ is $E_0$.

\section{Experimental Setup}\label{sec:experimental}

\begin{figure}[ht!]
  \centering\includegraphics[width=\linewidth]{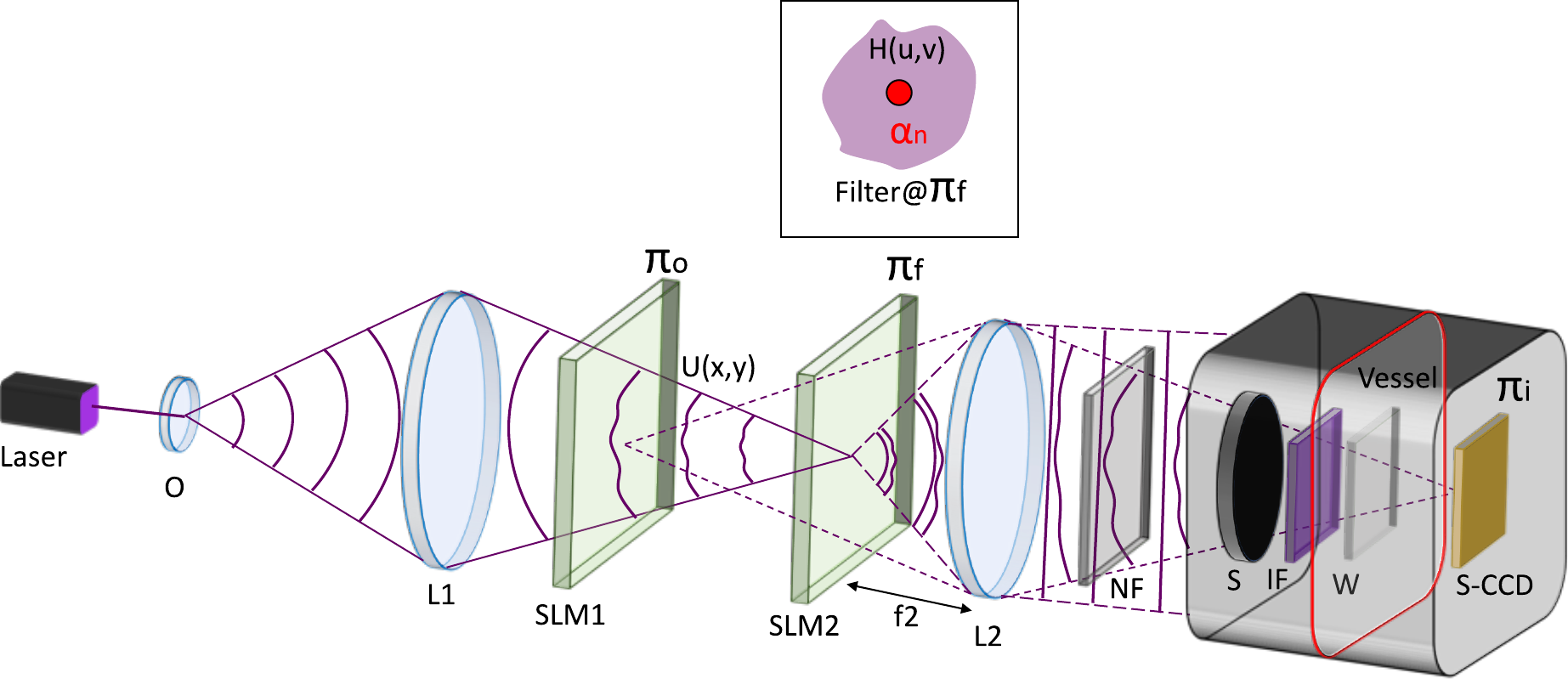}
  \caption{Experimental setup. The light source is a 405 nm cw laser diode,
    attenuated down to the single photon level. Lenses $Ls$ conform a
    convergent optical processor. SLMs are phase-only spatial light modulators.
    The interferograms are detected by a Skipper CCD. The detail shows the filter
    placed at the Fourier Plane $\Pi f$ where the central pixel of SLM2 introduces
    the PSI phase retardation $\alpha_n$.\label{fig:setup}}
  \end{figure}
The experimental setup, sketched in figure \ref{fig:setup}, is basically a
convergent optical processor, where two phase-only Spatial Light Modulators
(SLMs) are used: one to display the phase distribution to be measured $U(x, y)$,
and the other to dynamically introduce the phase retardation needed to implement
the PSI process. The light source is a laser diode at $405\ \mathrm{nm}$, that
is expanded by the microscope objective $O$ and spatially filtered by a pinhole
that is imaged by lens $L1$ onto the Fourier plane $\Pi f$. The phase
distribution $U(x, y)$ is displayed on the phase-only SLM1, which is placed at
the object plane $\Pi o$. When $U(x, y)$ is uniform, the resulting light
distribution on $\Pi f$ is a bright central spot, corresponding to the Fourier
transform of the entrance pupil of the system. At this position, a phase filter
$H(u, v)$, smaller than the focused spot, is displayed on SLM2. Then the phase
shifts $\alpha_n$ needed to implement the PSI technique are introduced in the
central pixel. (see inset in figure \ref{fig:setup}). This pixel is used as a
perturbation to generate a spherical wave by diffraction. When the phase
distribution $U(x, y)$ is not uniform, the bright spot is deformed. Most of the
light is diffracted towards higher spatial frequencies, and only the small
central part of the spot goes through the phase filter. After SLM2, the lens
$L2$ is placed at a distance equal to its focal length $f2$ and images the
object plane onto $\Pi i$. In this way, the spherical reference wavefront, which
is diffracted by the central pixel, is collimated and interferes with the tested
wavefront. Finally, the interferograms are detected by the Skipper-CCD placed at
the image plane $\Pi i$. The main advantage of this convergent configuration is
that it allows to change the size of the Fourier transform, to better suit the
size of the PDI filter.

Both SLMs are conformed by a Sony liquid crystal television panel model LCX012BL
which, in combination with polarizers and wave plates, that provide the adequate
state of light polarization, allows a $2\pi$ phase modulation
\cite{Marquez2001}. These liquid crystal displays have a VGA resolution of
640×480 and a pixel size of $43\ \mathrm{\mu m}$. Although the Skipper-CCD was
hitherto only used as a particle detector \cite{SENSEI2020, DAMICM2020,
  CONNIE2021, Rodrigues2021, SENSEI2022}, in a recent paper
\cite{PearsStefano2023} it was used for imaging purposes. This device offers
ultra-low readout noise and photon-number resolving capability. These features
are achieved by eliminating the low-frequency readout noise (around $2e^-$ in a
conventional CCD) by multiple, nondestructive measurements (samples) of the
charge in each pixel. For a readout noise of $0.2e^-$ (attained after 256
samples) the photon counting probability of misclassification is lower than
$1\%$ and can be further reduced just by increasing the number of charge samples
\cite{Rodrigues2021}. This technology also provides the lowest dark current
($\sim 10^{-4}\ \mathrm{e^-/pixel/day}$) \cite{SENSEI2022} and a full-well
capacity above 33000 electrons \cite{Drlica2020} in a spatially resolved sensor
with only (15x15) $\mu m$ pixel size. The Skipper-CCD sensor has a total active
area of 4126x886 pixels. It was designed at Lawrence Berkeley National
Laboratory and manufactured by Teledyne/DALSA using high-resistivity ($>10\mathrm{k}\Omega\cdot \mathrm{cm}$)
silicon wafers. As it operates in the temperature range of 135 to 140 K, in vacuum, the
sensor is placed into a vessel that has a fused silica window (W). A dark
chamber composed of a black-painted metallic holder was firmly attached to the
Skipper-CCD vessel. This holder is used to place an electronic shutter (S)
Melles Griot, that allows to control light exposure, and a square interferential
filter (IF) centered at the laser wavelength, that avoid spurious ambient light
from entering the sensor. Additionally, neutral filters (NF) are employed in
order to reduce the light intensity.

After the exposition to the light for each of these phase steps, the reading
process begins. This can be done in an efficient way, by selecting the amount of
charge samples NSAMP desired for each region of the sensor, thus optimizating the
total readout time of the measurement. The raw outcome is data in Analogical to
Digital Units (ADUs) for each pixel that needs to be calibrated to number of
electrons. Since the readout noise can be reduced to sub-electron values for
high NSAMP, there is a closed relation between this two variables with very low
probability of misclassification. The complete process for this calibration is
described in \cite{Rodrigues2021, PearsStefano2023}.

\begin{figure}[ht!]
\centering\includegraphics[width=\linewidth]{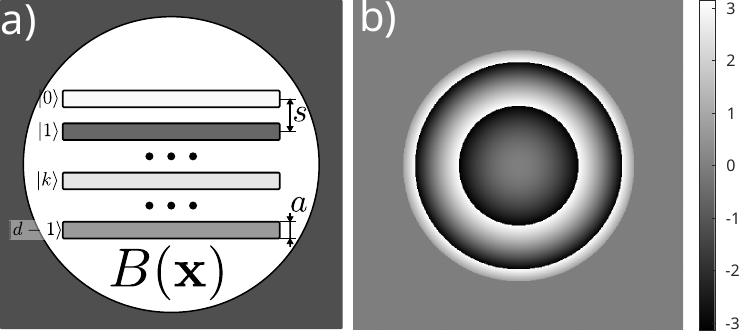}
\caption{Programmed phases in the first SLM. a) Slit photonic qudit state case; 
b) Continous lens-like phase map\label{fig:slm-phases}}
\end{figure}

\begin{figure*}[ht!]
  \centering\includegraphics[width=\linewidth]{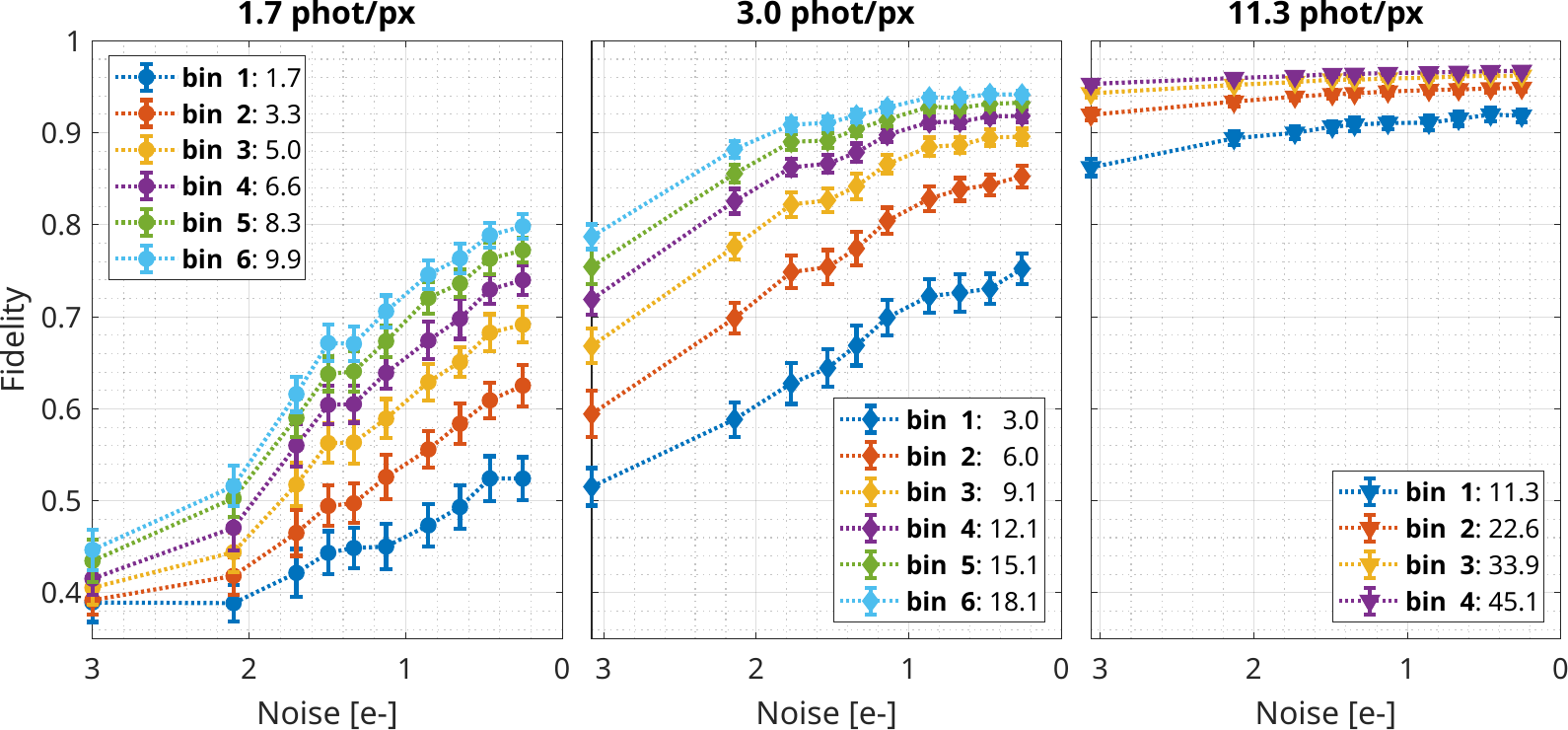}
  \caption{Reconstruction fidelity as a function of readout noise, for three
    different effective illuminations. a) $1.7\ \mathrm{phot/px}$; b)
    $3.0\ \mathrm{phot/px}$; c) $11\ \mathrm{phot/px}$; The different curves
    represent effective illuminations, where the state was estimated by
    selecting at random $n$ pixels for each slit.\label{fig:fidelity-noise}}
\end{figure*}

\begin{figure}[ht!]
\centering\includegraphics[width=\linewidth]{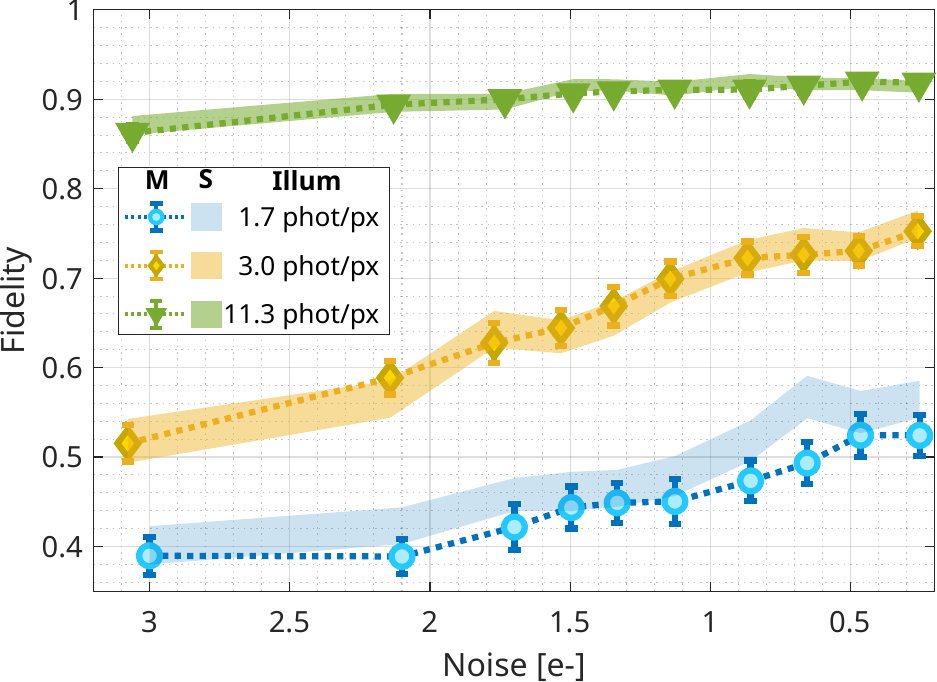}
\caption{Measured fidelities (errorbar) and simulated (points) as a function of
  readout noise for three different illuminations. The errorbars represent the
  standard error over 100 realizations of the selection of the pixels in the
  region of interest of each slit .\label{fig:experiment-simulation}}
\end{figure}

\begin{figure}[ht!]
\centering\includegraphics[width=\linewidth]{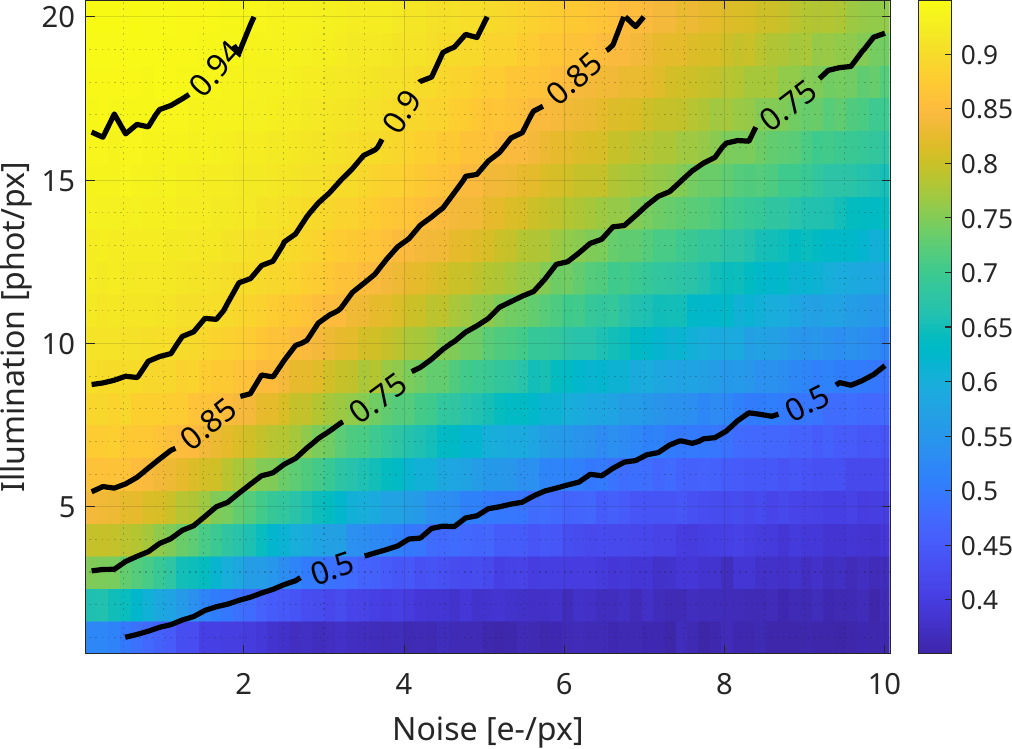}
\caption{Simulated map of mean fidelity as a function of the total illumination
  and the reading noise.\label{fig:simulation-map}}
\end{figure}

\section{Numerical simulations and experimental results}\label{sec:results}
At very low light levels two independent problems arise. On one hand the
fluctuations in number of photons becomes comparable to the average number of
photons, so that, the noise coming from the light source statistics affects the
accuracy of the phase estimation. On the other hand, the average number of
photons is itself comparable to the readout noise of the camera used to acquire
the interferograms. Thus, it is of interest to study, under these conditions,
what is the average minimum number of photons per pixel that is necessary to
carry out the phase reconstruction process with a certain degree of reliability.
And furthermore, how this number depends on the readout noise of the camera.

In this work we explored two cases of interest: the discrete phase representing
a qudit in the codification of \emph{slit states}, and a continuous quadratic
phase representing convergent lens.

\subsection{Slits states}
The first case is specially relevant for quantum information tasks. The
\emph{slits states}, whose formalism is described in detail in references
\cite{Neves2004,Solis-Prosser2013,Varga2017}, are a versatile codification that
harness the spatial degrees of freedom of a single photon, to allow the creation
of high dimensional quantum states. Particularly, the discretized transverse
momentum of single photons defines photonic quantum states. In a previous work
we showed\cite{PearsStefano2020} how to characterize these \emph{slits states}
with a PDI interferometric configuration.

Figure \ref{fig:slm-phases}a shows schematically the phase distribution that
represent the slits state. They consist on a phase mask with $d$ rectangular
slits of width $a$ and length $L (\gg a,s)$, where the separation between
adjacent slits is $s$. Thus, the resulting state is
\begin{align}\label{eq:qudit}
  |\Psi\rangle = \sum_{k=0}^{d-1} c_k | k\rangle, 
\end{align}
where $\left\{| k\rangle\right\}_{k=0}^{d-1}$ is the logical basis, and the
complex coefficients $c_k$ represent the transmissivity and phase of each of the
slits. Additionally, in order to have a light distribution similar to those
usually present in an in-line holography process (small diffracting objects
immersed in a strong background), we embed these slits in an constant light
background with uniform phase\cite{PearsStefano2020}.

As in this work we are focusing in phase determination, we choose a particular
state with uniform intensity in each slit, of dimension $d=6$, and with a phase
that increases in equal step between adjacent slits:
\begin{align}\label{eq:qudit-particular}
  |\Psi\rangle = \frac{1}{\sqrt{6}} \sum_{k=0}^{d-1} \exp(i 2 \pi k/5) | k\rangle.
\end{align} Thus, our target state contains optical phases in all the range between 0 and $2\pi$.

Once the phase of each pixel is estimated by Equation \ref{eq:phase}, as
described in Sec. \ref{sec:framework}, the state is reconstructed by at
random one pixel corresponding to each of the slits, and assigning that phase
value to that element of the canonical basis. As a figure of merit we used the
fidelity, that is defined, for pure states, as
$F(\Phi, \Psi) = |\langle\Phi|\Psi\rangle|$, where $|\Phi\rangle$ represents the
state to be prepared, and $|\Psi\rangle$ the state that is reconstructed
\cite{Nielsen}.

We reconstructed this state for three levels of illumination:
$1.7\ \mathrm{phot/px},$ $3.0\ \mathrm{phot/px},$ and $11.3\ \mathrm{phot/px}.$
Where the illumination was estimated as the intensity averaged over all the
pixels in the region defined by the slits ($ \sim 600$). However, in order to study
other values of illumination, we considered the $n-$pixel (binning of $n$
pixels) averaged state, where the state determination was made by averaging the
phase for $n$ randomly chosen pixels in the region of each of the slits. In this
case we defined the effective illumination as being $n$ times the real
illumination per pixel. To estimate the mean value of the fidelity, and its
dispersion in each case, we perform a boostrap method: for each level of noise
and binning $n$, we sampled without repetition 81 states, to get an average
fidelity. This process was repeated 64 times to obtain the mean value of these averaged
fidelities, and its standard deviation.

Figure \ref{fig:fidelity-noise} shows how the reconstruction fidelity
changes with the readout noise, for the three cases of real illumination. Each
of the curves in a single plot, represent the effective illumination, that was
obtained by binning $n$ pixel for each of the slit regions, to determine the
phase of said slit. It is clear from these plots the improvement in the fidelity
attained by increasing the number of samples in the Skipper CCD mode. The
highest readout noise corresponds to $\mathrm{NSAMP}=1$ samples in the Skipper CCD
($\sim 3e^-$), while the smallest readout noise corresponds to $\mathrm{NSAMP} = 144$ samples
($\sim 0.2e^-$).

In all cases the reduction in the readout noise leads to an increase on the
fidelity reconstruction. However, for the lowest level of illumination,
$1.7\ \mathrm{phot/px},$ maximum achieved fidelity for 1 bin is about $.5$,
pointing that this number of photons is not sufficient to determine the phase.
The fidelity in this case can be improved by averaging the phase over more
pixels, and thus incrementing the effective illumination. However, there is a
striking difference with the other two levels of illumination. For the
$3.0\ \mathrm{phot/px},$ the 1 bin case can achieve more than $.75$ of fidelity
when readout noise is decreased well below one $e^-$. And a binning greater or
equal to 2 is enough to get mean reconstruction fidelities over $.8$, for the
same readout noise condition. In the last case, with over
$11\ \mathrm{phot/px},$, the worst scenario corresponds to a fidelity that is  already over $.85$,
and the improvements are smaller both with the increased binning and the
decrease in the readout noise. This is expected, both the relative effect of
Poissonian fluctuations and that of the readout noise are much smaller.

\subsubsection{Simulation}
To perform the simulation, for each configuration of light level, synthetic
interferograms were generated according to the equation \ref{eq:en}. To account
for the Poissonian noise characteristic of the photon statistic of a highly
attenuated coherent state, we sampled, for each of the pixels of each of the
interferograms, a Poissonian distribution with the mean value corresponding to
the intensity value of said pixel. And additional Gaussian noise is added to 
each pixel, with zero mean and standard deviation equal to the electronic noise, 
to represent the readout noise of the camera.

Figure \ref{fig:experiment-simulation} compares the measured and the simulated
fidelity reconstruction as a function of the readout noise of the camera, for
three different examples of illumination: $1.7\ \mathrm{phot/px},$
$3.0\ \mathrm{phot/px},$ and $11.3\ \mathrm{phot/px}.$ In this case, the binning 
was $n=1$, and the random sampled was repeated 100 times. The continuous line 
represents  he average fidelity over the repetitions, while the error bars represent the 
standard error. On the other hand, the point represents the simulated reconstruction 
fidelity, averaged over 2000 repetitions. For the two highest illuminations, the noise model is in
excellent agreement with the measurements. In the case of the lowest illumination 
level, the simulated mean fidelity it is consistently above the value estimated 
experimentally, but the confidence band overlap for most of the points.

Thus, this noise model is a useful tool to estimate what is the expected reconstruction 
fidelity, given the level of illumination and the readout noise of the detector used. In that 
regard, Figure \ref{fig:simulation-map} shows 
the mean fidelity map as a function of the electronic
noise and the mean illumination, where in each point is an average over 2000
repetitions. To asses this fidelity for another dimension, the similar map can
be generated using the same statistical model.

\begin{figure*}[ht!]
\centering\includegraphics[width=\linewidth]{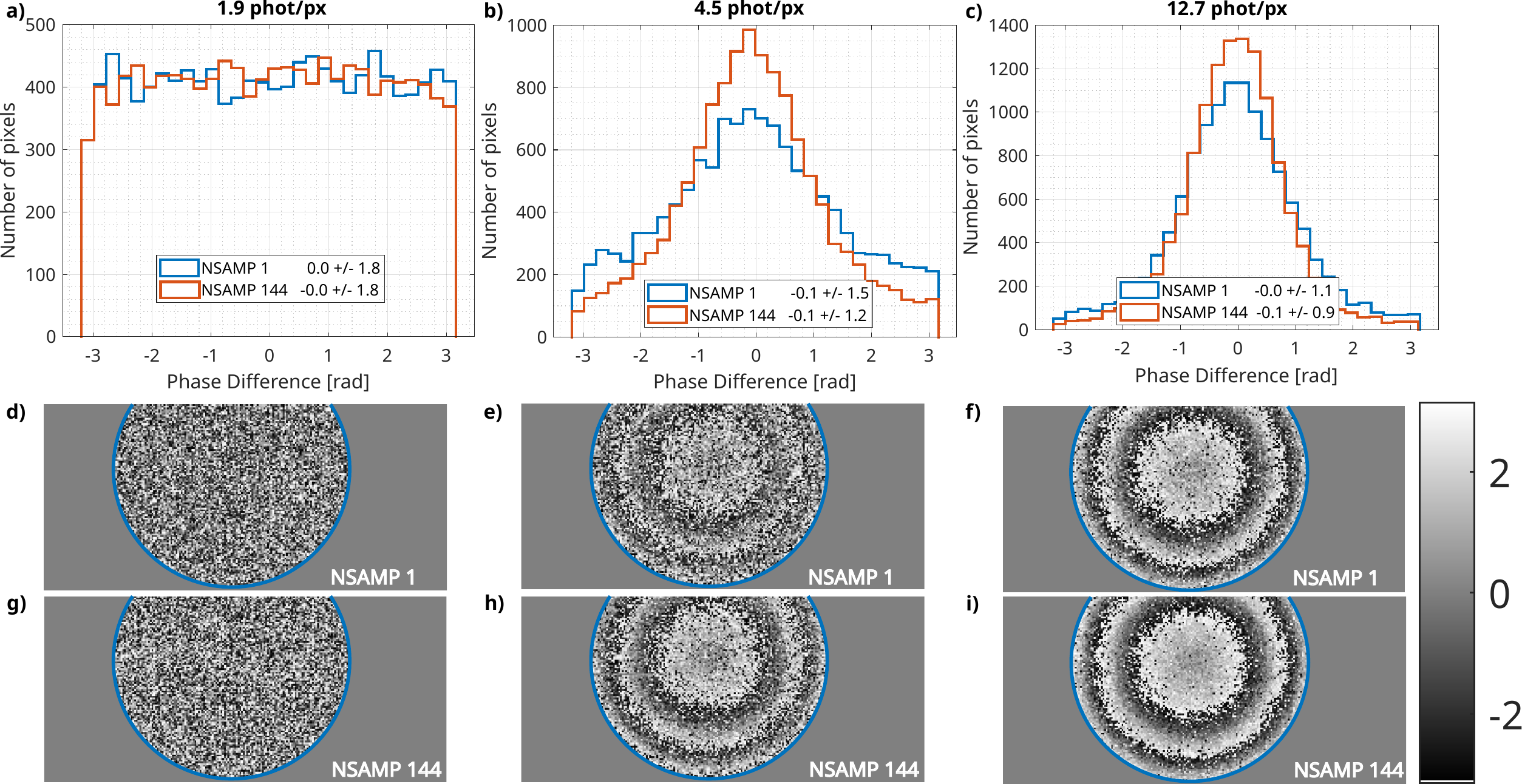}
\caption{Each column shows, for a fixed illumtion intensity, the histogram of phase difference 
in each pixel for a lens-like phase wavefront (top row), the corresponding phase map 
for NSAMP$=1$ (middle row), and for NSAMP=$144$ (bottom row). Panels a,d,g) $1.9\ \mathrm{phot/px}$; 
b,e,h) $4.0\ \mathrm{phot/px}$; c,f,i) $12.7\ \mathrm{phot/px}$;\label{fig:continous-phase}}
\end{figure*}

It should be noted that our analysis is based on raw images, that is, we do not
use maximum likelihood algorithms \cite{Barmherzig2022} to improve them, nor do
we apply principles of image compression and associated image reconstruction
\cite{Morris2015}. We also do not improve the sensitivity of the measurement by
using photon-subtracted thermal states \cite{HashemiRafsanjani2017}.

\subsection{Measurement of an arbitrary phase distribution}
Another case of interest is that of the determination of an arbitrary phase distribution. In
order to study this case, we programmed in the SLM1 a lens-like phase, i.e., a
quadratic phase, as represented schematically in Figure \ref{fig:slm-phases}b.

As in this case we can not define regions of constant phase, to asses the
quality of the phase reconstruction, we compare the obtained phase map with a
reference one, that was obtained with high light intensity
($\sim 500\ \mathrm{phot/px}$, where both the Poissoinan and the readout noise
are negligible).

The upper row of Figure \ref{fig:continous-phase} shows the histogram of the
phase differences for three different illuminations: $1.9\ \mathrm{phot/px},$
$4.5\ \mathrm{phot/px},$ and $12.7\ \mathrm{phot/px},$ and the two lines in each
histogram corresponds to the measurement without Skipper CCD mode
($\mathrm{NSAMP}=1$) and with ($\mathrm{NSAMP}=144$). The lower rows show a
comparison between the reconstructed phase maps for the minimum ($\mathrm{NSAMP}=1$, medium row) and the
maximum number ($\mathrm{NSAMP}=144$, bottom row) of Skipper CCD samples that we used. It is
worth noting that the quality of the reconstruction for $1.9\ \mathrm{phot/px},$
(Figure \ref{fig:continous-phase}a) does not improve even when reducing the
readout noise well below the one $e^-$ level. This is consistent with the
results that we got in the discrete case (Figure \ref{fig:fidelity-noise}a - bin
1), where with approximately the same level of illumination, the reconstruction
fidelity could not be increased more \mbox{than $\sim .5$}. 

Instead, for the cases of Figures \ref{fig:continous-phase}b and \ref{fig:continous-phase}c, 
there is a improvement in the phase estimation with the number of Skipper CCD samples, 
reflected in a decrease of $\sim 20\ \%$ in the standard deviation of the phase difference. 
In these two cases, a clear improvement in the phase map can be seen as the readout noise 
is reduced. However, for the $12.7\ \mathrm{phot/px}$ case, the improvement is smaller, 
as both the readout noise and the Poissonian noise start to be significantly lower than the mean intensity.

Thus, the phase determination can be significantly improved by reducing 
the readout noise, by taking advantage of the Skipper mode, for medium illumination 
levels, between 4 and 15 photons per pixel approximately. Interestingly, even when 
the readout noise is negligible, as is the case with the Skipper CCD camera, still a
 light level slightly higher than 4 photons per pixels is needed to reconstruct phases.  

\section{Conclusions}\label{sec:conclusions}
In this work we studied how interferometric phase reconstruction is affected by the 
two main sources of noise for highly attenuated coherent illumination: the readout 
noise of the detector used to record the interferograms, and the shot noise of the illumination. 

We reconstructed the phase of several wavefronts with a combination of a point 
diffraction interferometer and phase shifting interferometry, where the detection 
was done by Skipper-CCD camera. Allowing us to select, and even eliminate the readout noise.

On one hand we performed the reconstruction of a spatial distribution with $d=6$ 
uniform phase regions that encode a $d-$dimensional qudit. In this case, we showed an
 improvement in the reconstruction fidelity with the reduction of the readout noise for 
 all the illumination levels. However, even with in the absence of readout noise, with 
 under $3\ \mathrm{phot/px}$ it was not possible to reconstruct states with a fidelity 
 over $.8$. In this particular case, we propose a noise model to predict the  expected
  fidelity reconstruction as a function of both the readout noise level, and the mean 
  number of photons, that is consistent with the experimental results.

On the other hand, we reconstructed a continuous phase, showcasing phase 
reconstruction under low illumination for arbitrary phases. In a similar 
manner to the discrete case, the quality of the reconstruction was improved by
 the reduction of the readout noise for illuminations greater than $4\ \mathrm{phot/px}$.
  And likewise, even in the absence of readout noise, non meaningful phase could be 
  estimated with less than $3\ \mathrm{phot/px}$.

\section*{Acknowledgments}

This work was supported by Fermilab under DOE Contract No.\ DE-AC02-07CH11359.
The CCD development work was supported in part by the Director, Office of
Science, of the DOE under No.~DE-AC02-05CH11231. CI and QPS acknowledges the
support of the Secretaría de Ciencia y Técnica, Universidad de Buenos Aires
(20020170100564BA), of the Consejo Nacional de Investigaciones Científicas y
Técnicas (PIP 2330), and of the Ministerio de Ciencia Tecnologia e
Innovacion/Agencia Nacional de Promocion Cientifica y Tecnologica
(PICT-2020-SERIEA-02031). QPS was supported by a \mbox{CONICET} Fellowship.

%\nocite{*}

\bibliography{bibliography}% 

\end{document}